\documentclass{llncs}
\usepackage{graphicx}
\DeclareGraphicsExtensions{.pdf,.png,.jpg}
\usepackage{amsfonts}
\usepackage[fleqn]{amsmath}
\usepackage{amssymb}
\usepackage{tikz}
\usetikzlibrary{decorations.markings} 
\usepackage[T1]{fontenc}
\usepackage{scalefnt}
\usepackage{makeidx}
\usepackage{algorithm}
\usepackage{xspace}
\usepackage{hyperref} 

\usetikzlibrary{petri}
\usetikzlibrary{positioning}
\usetikzlibrary{backgrounds}
\tikzset{
LTS/.style={node distance=3em, on grid},
state/.style={circle,fill=black,inner sep=0.05cm},
edge/.style={-latex, auto},
}

\usepackage{filecontents} 
\tikzset{
    pos plot label/.style 2 args={
        postaction={
            decorate,
            decoration={
                markings,
                mark=at position #1 with \node#2;
            }
        }
    }
}
\makeatletter
\tikzset{
    scale plot marks/.is choice,
    scale plot marks/false/.code={
        \def\pgfuseplotmark##1{\pgftransformresetnontranslations\csname pgf@plot@mark@##1\endcsname}
    },
    scale plot marks/true/.style={},
    scale plot marks/.default=true
}
\makeatother

\tikzset{
    LTSArc/.style={-Latex},
    PNFlow/.style={-Triangle}
}

\makeindex

\begin{document}

\newcommand{\lts}{{\rm LTS}\xspace}
\newcommand{\es}{\emptyset}
\DeclareRobustCommand{\pminus}{\mathbin{\ooalign{\hfil $-$\hfil \cr \hfil\raisebox{1.5mm}{$\scriptstyle\bullet$}\hfil}}} 
\newcommand{\leer}{\varepsilon}
\newcommand{\N}{\mathbb{N}}
\newcommand{\Z}{\mathbb{Z}}
\newcommand{\Q}{\mathbb{Q}}
\newcommand{\E}{\mathbb{E}}
\newcommand{\minus}{\setminus}
\newcommand{\emptyseq}{\varepsilon}
\newcommand{\impl}{\Rightarrow}
\newcommand{\parikh}{{\mathcal P}}
\def\tp{^{\sf T}}
\newcommand{\tile}[4]{\begin{array}{|ll|}\hline#1&#2\\#3&#4\\\hline\end{array}}
\newcommand{\support}{\mathit{supp}}
\newcommand{\one}{\mathbf{1}}
\newcommand{\zero}{\mathbf{0}}
\newcommand{\step}[1]{[#1\rangle}
\newcommand{\M}{{\mathcal M}}
\newcommand{\comment}[1]{\hfill $\triangleright$ #1}
\newcounter{mylinenr}[algorithm]
\newlength{\myindent}
\newcommand{\linenr}[1][]{\stepcounter{mylinenr}\addtolength{\myindent}{#1em}\makebox[1em][r]{\scriptsize\arabic{mylinenr}:}\hspace*{\myindent}}
\newcommand{\NP}{{\sf NP}}
\newcommand{\PP}{{\sf P}}

\newcommand{\disjcup}{\makebox[1.0em][c]{
\setlength{\unitlength}{0.4em}
\begin{picture}(2,2)(0,0)
\put(0.84,0.48){{\tiny $\bullet$}}
\put(0.5,0){$\cup$}
\end{picture}
\setlength{\unitlength}{1mm}
}}

\newcommand{\EndSy}{\hfill\protect\makebox[1.0em][c]{
\protect\setlength{\unitlength}{0.2em}
\protect\begin{picture}(3,3)(0,0)
        \begin{thinlines}
\protect\put(0,0){\line(1,0){3}}
\protect\put(0,0){\line(0,1){3}}
\protect\put(0,3){\line(1,0){3}}
\protect\put(3,0){\line(0,1){3}}
        \end{thinlines}
\protect\end{picture}
\protect\setlength{\unitlength}{1mm}
}}

\newcommand{\choice}{\protect\makebox[1.0em][c]{
\protect\setlength{\unitlength}{0.2em}
\protect\begin{picture}(2,4)(0,0)
\protect\put(0,0){\line(1,0){2}}
\protect\put(0,0){\line(0,0){4}}
\protect\put(0,4){\line(1,0){2}}
\protect\put(2,0){\line(0,1){4}}
\protect\end{picture}
\protect\setlength{\unitlength}{1mm}
}}

\newcommand{\BX}[1]{{\unskip\nobreak\hfil\penalty50
                    \hskip2em\hbox{}\hfil
\EndSy\/ {{\rm #1}}
                    \parfillskip=0pt \finalhyphendemerits=0 \par
                   }}

\newcommand{\DEF}[2]{\goodbreak\begin{definition}
                     \label{#1}\begin{rm}{\sc #2}\nopagebreak 

                    }
\newcommand{\ENDDEF}[1]{\BX{\ref{#1}}
                        \end{rm}\end{definition}
                       }
\newcommand{\ENXDEF}{\end{rm}\end{definition}}
\newcommand{\KRYPT}[2]{\goodbreak\begin{kryptosystem}
                     \label{#1}\begin{rm}{\sc #2}\nopagebreak 

                    }
\newcommand{\ENDKRYPT}[1]{\BX{\ref{#1}}
                        \end{rm}\end{kryptosystem}
                       }
\newcommand{\KOR}[2]{\goodbreak\begin{corollary}
                     \label{#1}{\sc #2}\nopagebreak 

                    }
\newcommand{\ENDKOR}[1]{\BX{\ref{#1}}
\end{corollary}
                       }
\newcommand{\ENXKOR}{
\end{corollary}}
\newtheorem{notation}[theorem]{{\bf Notation}} 
\newcommand{\PROP}[2]{\goodbreak\begin{proposition} 
                      \label{#1}{\sc #2}\nopagebreak 

                     }
\newcommand{\ENDPROP}{
\end{proposition}} 
\newcommand{\ENXPROP}[1]{\BX{\ref{#1}}
\end{proposition}} 
\newcommand{\THEO}[2]{\goodbreak\begin{theorem}
                     \label{#1}{\sc #2}\nopagebreak 

                    }
\newcommand{\ENDTHEO}{
\end{theorem}}
\newcommand{\SATZ}[2]{\goodbreak\begin{theorem}
                     \label{#1}{\sc #2}\nopagebreak 

                    }
\newcommand{\ENDSATZ}{
\end{theorem}}
\newcommand{\ENXSATZ}[1]{\BX{\ref{#1}}
\end{theorem}}
\newcommand{\LEM}[2]{\goodbreak\begin{lemma}
                     \label{#1}{\sc #2}\nopagebreak 

                    }
\newcommand{\ENDLEM}{
\end{lemma}}
\newcommand{\ENXLEM}[1]{\BX{\ref{#1}}
\end{lemma}}
\newcommand{\BEW}{\goodbreak
{\bf Proof:}
                 }
\newcommand{\XBEW}{\goodbreak{\bf Proof:} }
\newcommand{\ENDBEW}[1]{\BX{\ref{#1}}
                       }
\newcommand{\ENXBEW}{}
\newcommand{\BBEW}[1]{\goodbreak{\bf Beweis von {\rm #1}:}}

\newcommand{\ENDBBEW}[1]{\BX{\ref{#1}}
                        }
\newcommand{\ENXBBEW}{}
\newcommand{\NOT}[2]{\goodbreak\begin{notation}
                     \label{#1}\begin{rm}{\sc #2}\nopagebreak 

                    }
\newcommand{\ENDNOT}[1]{\BX{\ref{#1}}
                        \end{rm}\end{notation}
                       }
\newcommand{\ENXNOT}{\end{rm}\end{notation}}
\newcommand{\REM}[2]{\goodbreak\begin{remark}
                     \label{#1}\begin{rm}{\sc #2}
                    }
\newcommand{\ENDREM}[1]{\BX{\ref{#1}}
                        \end{rm}\end{remark}
                       }
\newcommand{\ENXREM}{\end{rm}\end{remark}}
\newcommand{\BSP}[2]{\goodbreak\begin{beispiel}
                     \label{#1}\begin{rm}{\sc #2}\nopagebreak 

                    }
\newcommand{\ENDBSP}[1]{\BX{\ref{#1}}
                        \end{rm}\end{bespiel}
                       }

\newcommand{\prefix}{\sqsubseteq}
\newcommand{\backwd}[1]{\stackrel{\leftharpoonup}{#1}}
\newcommand{\down}{\;\downarrow\!}
\newcommand{\uniqueP}{\Upsilon}
\newcommand{\vzero}{0}
\newcommand{\from}{\leftarrow}
\newcommand{\drop}[1]{}

\renewcommand{\labelitemi}{$\bullet$}
\renewcommand{\labelitemii}{$\circ$}
\renewcommand{\labelitemiii}{$\cdot$}
\renewcommand{\labelitemiv}{$\ast$}

\newif\ifcomment
\commenttrue 
\definecolor{gris}{gray}{0.3}
\newcommand{\mycomment}[3]{\ifcomment
 {\small \null\-
  {\color{#1}{\textbf{#2:}} \color{#1} {#3}}}%
 \fi
}\newcommand{\EB}[1]{\mycomment{red!80!black}{EB}{#1}}
\newcommand{\RD}[1]{\mycomment{green!60!black}{RD}{#1}}
\newcommand{\US}[1]{\mycomment{blue!60!black}{US}{#1}}
\newcommand{\HW}[1]{\mycomment{green!60!black}{HW}{#1}}

\newcounter{exampleTScounter}
\newcommand{\TSref}[1]{\TS_{\ref{#1}}}
\newcommand{\PNref}[1]{\mathit{PNS}_{\ref{#1}}}
\newcommand{\PNS}{\mathit{PNS}}
\newcommand{\Nref}[1]{N_{\ref{#1}}}

\newcommand{\np}{\mathbb{N}_{\neq 0}}
\newcommand{\qp}{\mathbb{Q}_{>0}}
\newcommand{\brief}{cycle-reduced}
\newcommand{\briefness}{cycle-reduction}

\renewcommand{\mod}{\text{ {\rm mod} }}

\itemsep0pt

\title{Optimal Label Splitting for Embedding an LTS into an arbitrary Petri Net Reachability Graph is NP-complete}

\author{Uli Schlachter\inst{1,2}\thanks{Supported by DFG (German Research Foundation)
through grant Be 1267/15-1 {\tt ARS} (Algorithms for Reengineering and Synthesis)} and
Harro Wimmel\inst{2}\thanks{Supported by DFG (German Research Foundation) 
through grant Be 1267/16-1 {\tt ASYST} (Algorithms for Synthesis and Pre-Synthesis Based on Petri Net Structure Theory).}
}

\institute{
Institute of Networked Energy Systems, \\
German Aerospace Center,
Oldenburg, Germany, \\
\email{uli.schlachter@dlr.de}
\and
Department of Computing Science,\\
Carl von Ossietzky Universit\"{a}t Oldenburg,
Oldenburg, Germany \\
\email{harro.wimmel@informatik.uni-oldenburg.de}
}

\maketitle

\begin{abstract}
For a given labelled transition system (\lts), synthesis is the task to find an unlabelled
Petri net with an isomorphic reachability graph. Even when just demanding an embedding into
a reachability graph instead of an isomorphism, a solution is not guaranteed. In such a case,
label splitting is an option, i.e. relabelling edges of the \lts such that differently labelled 
edges remain different. With an appropriate label splitting, we can always obtain a
solution for the synthesis or embedding problem. Using the label splitting, we can construct
a labelled Petri net with the intended bahaviour (e.g. embedding the given \lts in its reachability
graph). As the labelled Petri net can have a large number of transitions, an optimisation may be desired,
limiting the number of labels produced by the label splitting. We show that such a limitation
will turn the problem from being solvable in polynomial time into an NP-complete problem.
\end{abstract}


\section{Introduction}
\label{intro.sct}

There are two general approaches to investigate the behaviour of Petri nets \cite{reisig,murata-89}.
{\it Analysis} is used to construct a variety of descriptions from sets of
firing sequences~\cite{hack-lang} to event structures~\cite{npw81}. One of the most common forms for describing the
sequential behaviour is the reachability graph, containing the reachable markings 
as states together with edges denoting transitions that fire to reach 
one marking from another.
In the reverse direction, i.e.\ {\it synthesis}, we try to build an unlabelled Petri net that 
behaves like a given specification, e.g. a labelled transition system (\lts).
In Process Mining~\cite{vdaalst}, this can be used to find a small model covering a
large set of observable behaviours. While all those behaviours must be allowed
by the model, exact synthesis is often difficult to achieve. 
E.g., there is no unlabelled Petri net that produces the sequence $abbaa$ and nothing else.
We might overapproximate the behaviour by the \lts shown on the left of Fig.~\ref{f.gen0},
which can be synthesised into an unlabelled Petri net and embeds the sequence $abbaa$, but this 
approach is not always possible.
The right \lts of Fig.~\ref{f.gen0} is not embedded in any reachability graph. As $ab$ and $ba$ have
the same effect in any Petri net, the states $s_2$ and $s_5$ represent the same marking and will necessarily be
identified in its reachability graph.
Identifying the two states leads to additional behaviour, i.e.\ $abb$ and $baa$.

\begin{figure}[t]
\centering
\begin{tikzpicture}[LTS]
\node[state,label=below:$s_0$] (s0) {};
\path[edge] (s0) ++(-0.5,0) edge (s0);
\foreach \x/\y in {0/1,1/2,2/3,3/4,4/5} {
 \node[state,right=of s\x,label=below:$s_{\y}$] (s\y) {}; }
\node[state,above=of s2,label=above:$s_6$] (s6) {};
\draw[edge] (s0) edge node[auto]{$a$} (s1);
\draw[edge] (s1) edge node[auto]{$b$} (s2);
\draw[edge] (s2) edge node[auto]{$b$} (s3);
\draw[edge] (s3) edge node[auto]{$a$} (s4);
\draw[edge] (s4) edge node[auto]{$a$} (s5);
\draw[edge] (s2) edge node[auto]{$a$} (s6);
\end{tikzpicture}\hspace*{1cm}
\begin{tikzpicture}[LTS]
\node[state,label=right:$s_0$] (s0) {};
\path[edge] (s0) ++(-0.5,0) edge (s0);
\node[state,above right=of s0,label=above:$s_1$] (s1) {};
\node[state,right=of s1,label=above:$s_2$] (s2) {};
\node[state,right=of s2,label=above:$s_3$] (s3) {};
\node[state,below right=of s0,label=below:$s_4$] (s4) {};
\node[state,right=of s4,label=below:$s_5$] (s5) {};
\node[state,right=of s5,label=below:$s_6$] (s6) {};
\draw[edge] (s0) edge node[auto]{$a$} (s1);
\draw[edge] (s1) edge node[auto]{$b$} (s2);
\draw[edge] (s2) edge node[auto]{$a$} (s3);
\draw[edge] (s0) edge node[auto,swap]{$b$} (s4);
\draw[edge] (s4) edge node[auto]{$a$} (s5);
\draw[edge] (s5) edge node[auto]{$b$} (s6);
\end{tikzpicture}
\caption{Left: The sequence $abbaa$ is embedded in this \lts, which is the reachability graph of an
unlabelled Petri net. Right: An \lts that cannot be embedded into the reachability graph of any
unlabelled Petri net, as every reachability graph will identify the states $s_2$ and $s_5$ due to
$ab$ and $ba$ producing the same marking.}
\label{f.gen0}
\end{figure}
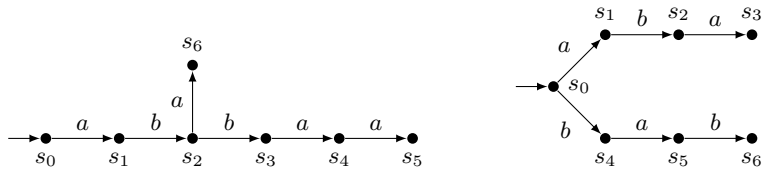

As an alternative, we may split the label $a$ into $a$ and $a'$, relabel e.g.\ $abbaa$ to $a'bbaa$, 
and find a Petri net for the new sequence $a'bbaa$.
With a labelling function $a,a'\mapsto a$ and $b\mapsto b$ we obtain a labelled Petri net with
the sought behaviour.
Label splitting~\cite{carmona-label-splitting,vanden-broucke-de-weerdt-2017,lu-fahland-etal-2016,sanpedro-cortadella-2016},
as indicated above, may lead to large Petri nets with as many transitions as there are edges in the \lts.
We hope to avoid this by allowing over-approximations of behaviour, i.e.\ embeddings into reachability graphs,
as well as by optimising the label splitting itself by limiting the number of newly introduced labels.
Essentially, we are interested in the following problem:

\begin{center}\framebox{\parbox{10cm}{
For a given \lts, is there a label splitting producing at most $n$ different labels such that the
relabelled \lts is embedded in the reachability graph of some (arbitrary) unlabelled Petri net?
}}\\[1mm]
or equivalently\\[1mm]
\framebox{\parbox{10cm}{
Can we find an arbitrary, {\it labelled} Petri net with at most $n$ transitions such that a given \lts
is embedded in its reachability graph?
}}\end{center}

An optimal solution can be determined by incrementing $n$ (starting at the alphabet size of the \lts)
until the problem is solved.
While, in general, synthesis and finding an embedding can be done in polynomial time using Region theory~\cite{ehren-roz-90,bbd},
we will show that finding an appropriate label splitting solving our problem is already \NP-complete.

In the next section, we introduce the basic concepts around labelled transition systems and
Petri nets as well as label splitting and a short description of synthesis. We also show that
the state separation property of Region theory, requiring different Petri net markings for different
states in the \lts, is equivalent to the existence of a reachability graph embedding the \lts.
Section~\ref{sect.NP} 
formalises our label splitting problem, shows membership in \NP, and provides the construction necessary to prove its
\NP-hardness. The construction is a generic \lts, consisting of six rather independent parts, the first four
of which are auxiliary in nature.
We take a closer look at these in Section~\ref{sect.units_and_region_values} and explain
how region values (i.e.\ marking changes) of labels
are determined. In Section~\ref{sect.polynomial_time} 
we show that our construction is a polynomial-time reduction, which proves
the \NP-completeness.
Finally, we give a summary and an outlook in Section~\ref{sect.O}.

\section{Basic concepts}
\label{basic-not.sct}

\DEF{d.lts}{LTS}
A {\it labelled transition system} (\lts) with initial state is a tuple $TS=(S,\Sigma,E,s_0)$
with nodes $S$ (a countable set of states),
edge labels $\Sigma$ (a finite set of letters),
edges $E\subseteq(S\times \Sigma\times S)$,
and an initial state $s_0\in S$. An edge $(s,t,s')\in E$ may be written as
$s\step{t}s'$.  
A {\it walk} $\sigma\in \Sigma^*$ from $s$ to $s'$, written as
$s\step{\sigma}s'$, 
is given inductively by $s=s'$ for the empty word $\sigma=\varepsilon$ and by
$\exists s''\in S$: $s\step{w}s''\step{t}s'$ for $\sigma=wt$ 
with $w\in \Sigma^*$ and $t\in \Sigma$. A walk $s\step{\sigma}s'$ is a {\it cycle} 
if and only if $s=s'$.
The {\it Parikh vector} $\parikh(\sigma)\colon\Sigma\to\Z$ of a word $\sigma\in \Sigma^*$ maps each letter $t\in \Sigma$ 
to its number of occurrences in $\sigma$, it will often be written as an element of the
group spanned by $\Sigma$. The neutral element is written as \(\zero\) and comparisons are done componentwise.
We map to $\Z$ here instead of $\N$ to be able to extend the notion of a Parikh vector later and
to handle differences of Parikh vectors more easily.

A {\it spanning tree} $\Theta$ of $TS$ is a set of edges $\Theta\subseteq E$ 
such that for every $s\in S$ there is a unique walk from $s_0$ to $s$ using edges in $\Theta$ only.
This implies that $\Theta$ is cycle-free.
A {\it walk in $\Theta$} is a walk that uses edges in $\Theta$ only (and not any of $E\backslash\Theta$).
Edges in $E\backslash \Theta$ are called {\it chords}.
The {\it Parikh vector of a state} $s$ in a spanning tree $\Theta$ is $\parikh_\Theta(s) = \parikh(\sigma)$
where $s_0\step{\sigma}s$ is the unique walk in $\Theta$. The {\it Parikh vector of an edge}
$s\step{t}s'$ in $TS$ is $\parikh_\Theta(s\step{t}s') = \parikh_\Theta(s)+1t-\parikh_\Theta(s')$, see
Fig.~\ref{f.gen1} for an example. 
Note that Parikh vectors of edges in $\Theta$ always evaluate to zero; for chords the Parikh vector may even 
contain negative values. For a chord $s\step{t}s'$, $s$ and $s'$ have a latest common predecessor $r$ 
in $E$, $t',t''\in T$ with $t'\not=t''$, $\sigma,\sigma'\in T^*$ with two walks
$r\step{t'\sigma}s\step{t}s'$ and $r\step{t''\sigma'}s'$ in $\Theta$. These two walks form a cycle in the
LTS' underlying undirected graph, called a {\it generalised cycle}, with the Parikh vector the same as the chord's, 
$\parikh_\Theta(s)+1t-\parikh_\Theta(s')$. 
The {\it Parikh vector of a walk} $s_1\step{t_1}s_2\ldots s_n\step{t_n}s_{n+1}$ is defined as 
$\parikh_\Theta(s_1\step{t_1\ldots t_n}s_{n+1}) = \sum_{i=1}^n\parikh_\Theta(s_i\step{t_i}s_{i+1})$.
Obviously, $\parikh_\Theta(s_1\step{t_1\ldots t_n}s_{n+1}) = \parikh_\Theta(s_1)+\parikh(t_1\ldots t_n)-\parikh_\Theta(s_{n+1})$.
If the walk is a cycle (with $s_1=s_{n+1}$), we thus find 
$\parikh(t_1\ldots t_n)=\sum_{i=1}^n\parikh_\Theta(s_i\step{t_i}s_{i+1})$ where all non-zero
Parikh vectors in the sum stem from chords. 
The set $\{\parikh_\Theta(s\step{t}s')\,|\,(s,t,s') \in E\backslash \Theta\}$ is then a generator
for all Parikh vectors of cycles (the latter being linear combinations of its elements).
By simple linear algebra, 
we can compute a basis from this generator. This {\it cycle base} $\Gamma$ contains
at most $|\Gamma|\le|\Sigma|$ different Parikh vectors.

\begin{figure}[t]
\centering
\begin{tikzpicture}[LTS]
\node[state,label=below:$s_0$] (s0) {};
\path[edge] (s0) ++(-0.5,0) edge (s0);
\foreach \x/\y in {0/2,2/4,4/6} {
 \node[state,above=of s\x,label=left:$s_{\y}$] (s\y) {}; }
\foreach \x/\y in {0/1,2/3,6/7} {
 \node[state,right=of s\x,label=right:$s_{\y}$] (s\y) {}; }
\draw[edge] (s0) edge node[auto,swap]{$a$} (s1);
\draw[edge] (s0) edge node[auto]{$b$} (s2);
\draw[edge] (s2) edge node[auto]{$b$} (s4);
\draw[edge] (s4) edge node[auto]{$b$} (s6);
\draw[edge] (s6) edge node[auto]{$a$} (s7);
\draw[edge] (s1) edge node[auto,swap]{$b$} (s3);
\draw[edge,dashed] (s3) edge node[auto,swap]{$c$} (s0);
\draw[edge,dashed] (s7) edge node[auto,swap]{$c$} (s4);
\end{tikzpicture}\hspace*{1cm}
\begin{tikzpicture}[LTS]
\node[state,label=below:$s_0$] (s0) {};
\path[edge] (s0) ++(-0.5,0) edge (s0);
\foreach \x/\y in {0/2,2/4,4/6} {
 \node[state,above=of s\x,label=left:$s_{\y}$] (s\y) {}; }
\foreach \x/\y in {0/1,2/3,4/5,6/7} {
 \node[state,right=of s\x,label=right:$s_{\y}$] (s\y) {}; }
\draw[edge] (s0) edge node[auto,swap]{$a$} (s1);
\draw[edge] (s0) edge node[auto]{$b$} (s2);
\draw[edge] (s2) edge node[auto]{$b$} (s4);
\draw[edge] (s4) edge node[auto]{$b$} (s6);
\draw[edge] (s6) edge node[auto]{$a$} (s7);
\draw[edge] (s1) edge node[auto,swap]{$b$} (s3);
\draw[edge,dashed] (s3) edge node[auto,swap]{$c$} (s0);
\draw[edge,dashed] (s7) edge node[auto,swap]{$c$} (s4);
\draw[edge,dashed] (s2) edge node[auto]{$a$} (s3);
\draw[edge] (s4) edge node[auto]{$a$} (s5);
\draw[edge,dashed] (s5) edge node[auto,swap]{$c$} (s2);
\end{tikzpicture}\hspace*{1cm}
\begin{tikzpicture}
\draw(0,0) node[transition] (a) {$a$};
\draw(2,2) node[transition] (c) {$c$};
\draw(2,0) node[place,tokens=5] (p1) {};
\draw(1,1) node[place,tokens=1] (p2) {};
\draw(0,2) node[place] (p3) {};
\draw(4,2) node[place] (p4) {};
\draw(4,0) node[transition] (b) {$b$};
\draw(c) edge[-latex,thick] (p2);
\draw(p2) edge[-latex,thick] (a);
\draw(a) edge[-latex,thick] (p3);
\draw(p3) edge[-latex,thick] (c);
\draw(b) edge[-latex,thick] (p4);
\draw(p4) edge[-latex,thick] (c);
\draw(c) edge[-latex,thick] node[auto]{$3$} (p1);
\draw(p1) edge[-latex,thick] node[auto]{$2$} (a);
\draw(p1) edge[-latex,thick,bend left=10] node[auto]{$3$} (b);
\draw(b) edge[-latex,thick,bend left=10] node[auto]{$2$} (p1);
\end{tikzpicture}
\caption{Two \lts (all solid and dashed edges count) with spanning trees $\Theta$ (solid edges only), the left spanning tree being unique, in the middle we have
a choice between the edges $s_1\step{b}s_3$ and $s_2\step{a}s_3$ for $\Theta$. States $s_2$, $s_3$, $s_4$, and $s_7$
have Parikh vectors $\parikh_\Theta(s_2)=1b$, $\parikh_\Theta(s_3)=1a+1b$,
$\parikh_\Theta(s_4)=2b$, and $\parikh_\Theta(s_7)=1a+3b$. This
yields $\parikh_\Theta(s_2\step{a}s_3)=\parikh_\Theta(s_2)+1a-\parikh_\Theta(s_3)=\zero$ 
and $\parikh_\Theta(s_7\step{c}s_4)=\parikh_\Theta(s_7)+1c-\parikh_\Theta(s_4)=1a+1b+1c$ for the chords
$s_2\step{a}s_3$ and $s_7\step{c}s_4$ in the middle \lts. The other chords $s_5\step{c}s_2$ and $s_3\step{c}s_0$
yield the same result as $s_7\step{c}s_4$. The cycle base $\Gamma$ contains the only
non-zero Parikh vector of a chord, $1a+1b+1c$.
The Petri net on the right can be synthesised
from the middle \lts, which in turn embeds the \lts on the left, making it PN-embeddable.
The left \lts is not PN-synthesisable due to unsolvable ESSPs $(s_2,a)$ and $(s_4,a)$.
}
\label{f.gen1}
\end{figure}
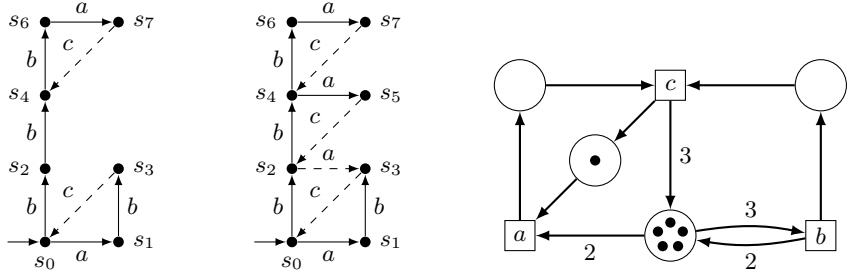

An LTS $TS=(S,\Sigma,E,s_0)$ is {\it finite} if $S$ is finite and it is {\it deterministic} 
if $s\step{t}s'\,\wedge\,s\step{t}s''$ implies $s'=s''$ for all $s\in S$ and $t\in \Sigma$. We call
$TS$ {\it reachable} if for every state $s\in S$ exists some $\sigma\in \Sigma^*$ with $s_0\step{\sigma}s$.
Reachability implies the existence of a spanning tree.
Two labelled transition systems $TS_1=(S_1, \Sigma_1, E_1, s_{01})$ and $TS_2=(S_2, \Sigma_2, E_2, s_{02})$ are 
{\it isomorphic} if $\Sigma_1=\Sigma_2$ and there is a bijection $\zeta\colon S_1\to S_2$ with $\zeta(s_{01})=s_{02}$ and
$(s,t,s')\in E_1 \Leftrightarrow(\zeta(s),t,\zeta(s'))\in E_2$, for all $s,s'\in S_1$. 
$TS_1$ can be {\it embedded} into $TS_2$ if $\Sigma_1\subseteq\Sigma_2$ and there is an injection 
$\zeta\colon S_1\to S_2$ with $s\neq s'$ implies $\zeta(s)\neq\zeta(s')$, such that 
$\zeta(s_{01})=s_{02}$ and
$(s,t,s')\in E_1 \Rightarrow(\zeta(s),t,\zeta(s'))\in E_2$ 
for all $s,s'\in S_1$.
\ENDDEF{d.lts}

Figure~\ref{f.gen1} shows an \lts an the left that can be embedded into the \lts in the middle
but is not isomorphic to it (due to $s_5$ and three additional edges). All shown edges, whether solid
or dashed, belong to the \lts in question. The solid edges present a spanning tree of the \lts.

\DEF{d.pn}{Petri nets}
An {\it (initially marked) Petri net} is denoted as $N=(P,T,W,M_0)$ where $P$ is a finite set of {\it places},
$T$ is a finite set of {\it transitions},
$W$ is the weight function $W\colon((P\times T)\cup(T\times P))\to\N$ 
specifying the {\it arc weights},
and $M_0$ is the {\it initial marking}
(where a {\it marking} is a mapping $M\colon P\to\N$, indicating the number of {\it tokens} in each place).
A transition $t\in T$ is {\it enabled} at a marking $M$,
denoted by $M\step{t}$, if $\forall p\in P\colon M(p)\geq W(p,t)$.
{\it Firing} $t$ leads from $M$ to $M'$, denoted by $M\step{t}M'$,
if $M\step{t}$ and $M'(p)=M(p)-W(p,t)+W(t,p)$.
This can be extended, by induction as usual, to $M\step{\sigma}M'$ for words $\sigma\in T^*$,
and $\step{M} = \{M'\,|\,\exists\sigma\in T^*\colon M\step{\sigma}M'\}$
denotes the set of markings reachable from $M$.
The {\it reachability graph} $RG(N)$ of a Petri net $N$
is the labelled transition system with the set of nodes
$\step{M_0}$,
initial state $M_0$, label set $T$,
and set of edges $\{(M,t,M')\mid M,M'\in\step{M_0}\land M\step{t}M'\}$.

If a labelled transition system $TS$ is isomorphic to the reachability graph $RG(N)$ of 
a Petri net $N$
we say that $N$ {\em PN-solves} (or simply {\em solves}) $TS$
or that $TS$ is {\em synthesisable} (to $N$). 
If $TS$ can be embedded into $RG(N)$,
we say that $N$ {\em over-approximates} $TS$
or that $TS$ is {\em PN-embeddable} (into $N$),
and write $TS\unlhd N$.
\ENDDEF{d.pn}

The \lts in the middle of Figure~\ref{f.gen1} is synthesisable to the Petri net on the right,
which can easily be seen by playing the token game, e.g. state $s_1$ corresponds to the marking
with one token on the upper left place and three on the lower one, which can be obtained by
firing $a$. The synthesisability of the middle \lts implies that the left \lts is PN-embeddable 
into this Petri net via the embedding between the two \lts.
If an LTS is not synthesisable or embeddable to any Petri net, we might opt to modify the LTS.
One way to do this is label splitting.

\DEF{d.split}{Label Splitting}
Let $TS=(S,\Sigma,E,s_0)$ be an LTS. A label splitting for $TS$ is a quadruple $(\Sigma',E',\varrho,\varphi)$
of a finite alphabet $\Sigma'\supseteq\Sigma$, a set $E'\subseteq S\times\Sigma'\times S$, a surjective mapping 
$\varrho\colon\Sigma'\to\Sigma$, and a bijection $\varphi\colon E'\to E$ 
such that $\varphi((s,t,s')\in E')=(s,\varrho(t),s')\in E$.
The result of the label splitting $(\Sigma',E',\varrho,\varphi)$ is the LTS $(S,\Sigma',E',s_0)$.

The label splitting $(\Sigma',E',\varrho,\varphi)$ is called {\em optimal} for $TS=(S,\Sigma,E,s_0)$
if $(S$, $\Sigma'$, $E'$, $s_0)$ is PN-embeddable (to an arbitrary Petri net) and every label splitting 
$(\Sigma'',E'',\varrho',\varphi')$ with a PN-embeddable LTS $(S,\Sigma'',E'',s_0)$ yields
$|\Sigma'|\le|\Sigma''|$.
\ENDDEF{d.split}

If an LTS $(S,\Sigma,E,s_0)$ is synthesisable or embeddable, 
an optimal label splitting is $(\Sigma,E,{\bf id},{\bf id})$. This is the case
for the two \lts in Fig.~\ref{f.gen1}. For the right \lts in Fig.~\ref{f.gen0} (written as $(S=\{s_0,\ldots,s_6\},\Sigma=\{a,b\},E,s_0)$),
which is not PN-embeddable, it is necessary to make the Parikh vectors of the paths
$s_0\step{ab}s_2$ and $s_0\step{ba}s_5$ distinguishable. One optimal label splitting
would be $(\{a,b,c\},E',\varrho,\varphi)$ with $\varrho(a)=\varrho(c)=a$, $\varrho(b)=b$,
$E'=\{(s_0,c,s_1),(s_1,b,s_2),(s_2,a,s_3),(s_0,b,s_4)$, $(s_4,a,s_5),(s_5,b,s_6)\}$,
$\varphi((s,t,s'))=(s,t,s')$ for $s,s'\in S$, $t\in\{a,b\}$, and 
$\varphi((s_0,c$, $s_1))=(s_0,\varrho(c),s_1)=(s_0,a,s_1)$.
Optimality of a label splitting leads to an over-approximating Petri net 
with a minimal number of transitions, in this case, three. The situation can become much more
complicated when the \lts contains cycles.

\DEF{d.synth}{Synthesis~\cite{bbd}}
A {\it region} $r=(R,B,F)$ of an LTS $(S,\Sigma,E,s_0)$ consists of three functions
$R$: $S\to\N$, $B$: $\Sigma\to\N$, and $F$: $\Sigma\to\N$ such that for all edges $s\step{t}s'$
in the LTS we have $R(s)\ge B(t)$ and $R(s')=R(s)-B(t)+F(t)$. 
The difference $\E(t)=F(t)-B(t)$ is called the {\it effect} of the label $t$. 
The defining conditions of a region mimic the firing
rule of Petri nets and make regions essentially equivalent to places, i.e.\ a place
$p$ can be defined from $r$ via $M_0(p)=R(s_0)$, $W(p,t)=B(t)$, and $W(t,p)=F(t)$ for
all $t\in \Sigma$. When a Petri net is constructed from a set of regions
of a reachable LTS, this implies
a {\it uniquely defined marking} $\M(s)$ for each state $s$ with 
$\M(s)(p)=M_0(p)+\sum_{t\in \Sigma}\parikh(\sigma)(t)\cdot(W(t,p)-W(p,t))$
for an arbitrary walk $s_0\step{\sigma}s$.

The construction of a Petri net $N=(P,T,W,M_0)$ with one place in $P$
for each region of the LTS guarantees $s\step{t} \impl \M(s)\step{t}$,
but has three issues: (1) $P$ might become infinite, (2) $\M$ may not be injective,
and (3) $\M(s)\step{t} \impl s\step{t}$ need not hold. Failing (2) means
that there are states $s,s'\in S$ with $s\not=s'$ that are identified in $RG(N)$ (leading to non-isomorphism).
A {\it state separation problem} (SSP) is a pair $(s,s')\in S\times S$ with $s\not=s'$.
A region $r$ solves an SSP $(s,s')$ if $R(s)\not=R(s')$ (and thus $\M(s)\not=\M(s')$).%
\footnote{An example for an SSP in Fig.~\ref{f.gen1} (both \lts) is $(s_3,s_7)$. Any region $r=(R,B,F)$ with $\E(b)\neq 0$ solves this SSP,
as this implies $R(s_3)=R(s_0)+\E(a)+\E(b)\neq R(s_0)+\E(a)+3\E(b)=R(s_7)$. We might choose $R(s_0)=0$,
$\E(a)=1=\E(b)$, and $\E(c)=-2$, for example, to obtain $R(s_3)=2\neq 4=R(s_7)$.}
Failing (3) results in an edge $\M(s)\step{t}$ in $RG(N)$ but not in the LTS,
$\neg s\step{t}$.
An {\it event/state separation problem} (ESSP) is a pair $(s,t)\in S\times \Sigma$
with $\neg s\step{t}$. A region $r$ solves an ESSP $(s,t)$ if $R(s)<B(t)$
(and thus $\neg\M(s)\step{t}$).%
\footnote{The left \lts of Fig.~\ref{f.gen1} has two unsolvable ESSPs, but ESSPs will not be important in this paper.}
The set of all {\it separation problems}, $\{(s,s')\in S\times S\,|\,s\not=s'\}\cup\{(s,t)\in S\times \Sigma \mid \neg s\step{t}\}$,  
is finite for finite LTS, and finding
a solution for every separation problem solves all three issues, making the synthesis successful with $RG(N)$
isomorphic to the finite LTS.
\ENDDEF{d.synth}

SSPs and ESSPs can be written as linear inequality systems over variables
$F\ge\zero$, $B\ge\zero$, and $R(s_0)\ge 0$. These systems are generally
polynomial in the size of the \lts, see e.g.~\cite{BDS-SOFSEM-2016,sw18}.

%

If the \lts is finite, the
linear inequality systems are also finite, and we may employ standard means, e.g.\ 
an ILP- or SMT-solver~\cite{smtinterpol}, to solve them. If an \lts is not reachable, it is not structurally isomorphic
to a reachability graph, i.e.\ no Petri net solving it exists.
The \lts may still be embeddable, but possibly into many different reachability graphs.
Depending on how the unreachable parts of the \lts will be connected after the embedding,
states must be mapped to other markings, so we may not be able to find a unique marking $\M(s)$ for every state $s$,
see Fig.~\ref{f.gen3} for an example. 
For these reasons, we assume all LTS to
be finite and reachable in the remainder of this paper.

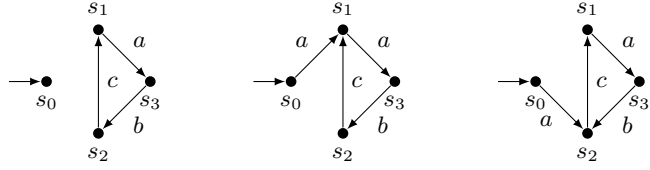
\begin{figure}[t]
\centering
\begin{tikzpicture}[LTS]
\node[state,label=below:$s_0$] (s0) {};
\path[edge] (s0) ++(-0.5,0) edge (s0);
\node[state,above right=of s0,label=above:$s_1$] (s1) {};
\node[state,below right=of s0,label=below:$s_2$] (s2) {};
\node[state,below right=of s1,label=below:$s_3$] (s3) {};
\draw[edge] (s1) edge node[auto]{$a$} (s3);
\draw[edge] (s3) edge node[auto]{$b$} (s2);
\draw[edge] (s2) edge node[auto,swap]{$c$} (s1);
\end{tikzpicture}\hspace*{1cm}
\begin{tikzpicture}[LTS]
\node[state,label=below:$s_0$] (s0) {};
\path[edge] (s0) ++(-0.5,0) edge (s0);
\node[state,above right=of s0,label=above:$s_1$] (s1) {};
\node[state,below right=of s0,label=below:$s_2$] (s2) {};
\node[state,below right=of s1,label=below:$s_3$] (s3) {};
\draw[edge] (s1) edge node[auto]{$a$} (s3);
\draw[edge] (s3) edge node[auto]{$b$} (s2);
\draw[edge] (s2) edge node[auto,swap]{$c$} (s1);
\draw[edge] (s0) edge node[auto]{$a$} (s1);
\end{tikzpicture}\hspace*{1cm}
\begin{tikzpicture}[LTS]
\node[state,label=below:$s_0$] (s0) {};
\path[edge] (s0) ++(-0.5,0) edge (s0);
\node[state,above right=of s0,label=above:$s_1$] (s1) {};
\node[state,below right=of s0,label=below:$s_2$] (s2) {};
\node[state,below right=of s1,label=below:$s_3$] (s3) {};
\draw[edge] (s1) edge node[auto]{$a$} (s3);
\draw[edge] (s3) edge node[auto]{$b$} (s2);
\draw[edge] (s2) edge node[auto,swap]{$c$} (s1);
\draw[edge] (s0) edge node[auto,swap]{$a$} (s2);
\end{tikzpicture}
\caption{The \lts with unreachable states on the left can be embedded into the other two \lts, which are
both synthesisable. The unique marking $\M$ for the left \lts (as far as it can be derived) must distinguish each pair of states, especially
the states $s_1$ and $s_2$, and determines the effect of $a$ as $\M(s_3)-\M(s_1)$.
Then, $\M(s_0)=\M(s_1)-(\M(s_3)-\M(s_1))$ must be different from $\M(s_0)=\M(s_2)-(\M(s_3)-\M(s_1))$,
showing that $\M(s_0)$ needs different values in the middle and right \lts.
This contradicts the uniqueness of $\M$ for the left \lts, i.e. we cannot distinguish the states in the left \lts without
knowing how they will be embedded}
\label{f.gen3}
\end{figure}

If we just aim at embedding an LTS into a Petri net, dealing with ESSPs is not necessary.
For elementary net systems this is already known~\cite{bbd}. 
For arbitrary Petri nets a similar result can be derived. 

\LEM{l.embedssp}{Embedding and SSP}
Let $TS=(S,\Sigma,E,s_0)$ be a finite, reachable LTS. $TS$ is PN-embeddable if and only if 
every SSP $(s,s')$ (with $s,s'\in S$, $s\neq s'$) is solvable.
\ENDLEM
\BEW
$\Rightarrow$: Let $(s,s')$ be an SSP and let $N=(P,T,W,M_0)$ be a Petri net
with $TS\unlhd N$ by an injective morphism $\zeta\colon S\to\step{M_0}$. If $s\neq s'$, we also
have $\zeta(s)\neq\zeta(s')$, i.e. there is a place $p\in P$ with $\zeta(s)(p)\neq\zeta(s')(p)$.
We find a region $r=(R,B,F)$ for $RG(N)$ with $R(s_0)=M_0(p)$ and $\forall t\in T$:
$B(t)=W(p,t)$ and $F(t)=W(t,p)$ such that $R(\zeta(s))\neq R(\zeta(s'))$. As $\zeta$ is
a morphism, $(\zeta\circ R,B,F)$ is a region for $TS$. 
Thus, the SSP $(s,s')$ is solved by the region $(\zeta\circ R,B,F)$.

$\Leftarrow$: Assume all SSPs $(s,s')$ are solvable. Each solution forms a region $r=(R,B,F)$,
giving rise to a place $p\in P$ with $M_0(p)=R(s_0)$, $W(p,t)=B(t)$, and $W(t,p)=F(t)$ for all $t\in\Sigma$.
Construct a Petri net $N=(P,\Sigma,W,M_0)$ with all these places.
For every state $s\in S$, we then obtain the uniquely defined marking $\M(s)$ with
$\M(s)(p)=M_0(p)+\sum_{t\in \Sigma}\parikh(\sigma)(t)\cdot(W(t,p)-W(p,t))$ for every place $p$ and
walk $s_0\step{\sigma}s$ in $TS$.
$\M$ is obviously injective, as two different states $s$ and $s'$ are distinguished at least by
the region solving the SSP $(s,s')$ and the place $p$ constructed from it.
Let $(s,t,s')\in E$, so $\M(s)(p)=M_0(p)+\sum_{t\in\Sigma}\parikh_\Theta(s)(t)\cdot(W(t,p)-W(p,t))
=R(s_0)+\sum_{t\in\Sigma}\parikh_\Theta(s)(t)\cdot\E(t)=R(s_0)+\E\tp\cdot\parikh_\Theta(s)=R(s)\ge B(t)$
and also $\M(s')(p)=R(s')=R(s)+\E(t)$ by the definition of the region $r=(R,B,F)$ corresponding with $p$.
Thus, $\M(s)\step{t}\M(s')$ in $RG(N)$ and with $\M(s_0)=M_0$ we conclude that $\M$ is an injective morphism
embedding $TS$ in $RG(N)$, i.e. $TS\unlhd N$.
\ENDBEW{l.embedssp}

\section{A Reduction for Near-Optimal Label Splittings}\label{sect.NP}

For finding an optimal label splitting, we must determine the minimal number of
labels required to make an \lts PN-embeddable, i.e.\ solve an optimisation problem. 
If we turn this number into an input parameter of our problem, we can convert the latter
into a decision problem:

\DEF{d.noptrel}{A Decision Problem}
Let $TS=(S,\Sigma,E,s_0)$ be a finite, reachable \lts and $q\in\N$ a number.
The \emph{near-optimal label splitting problem $(TS,q)$} is the question whether there exists 
a label splitting $(\Sigma',E',\varrho,\varphi)$ with $|\Sigma'|\le q$ such that
$TS'=(S,\Sigma',E',s_0)$ is PN-embeddable, i.e.\ all SSPs in it are solvable (cf. Lemma~\ref{l.embedssp}).
\ENDDEF{d.noptrel}

If the decision problem is decidable, the optimisation problem is also decidable
with only polynomial overhead. Take $TS$ and check for $|\Sigma|\le q\le |E|$
whether $(TS,q)$ can be positively decided, starting with $q=|\Sigma|$ and incrementing.
When we get a positive answer (which must happen with $q=|E|$ at the latest), 
we have found an optimal label splitting.
We need to make at most $|E|-|\Sigma|+1$ decisions, a number polynomial in the size of $TS$. 

%
%

\SATZ{s.optrelnp}{Near-optimal Label Splitting is in NP}
For finite, reachable \lts $TS=(S,\Sigma,E,s_0)$ and $q\in\N$,
the near-optimal label splitting problem $(TS,q)$ is in \NP.
\ENDSATZ
\BEW
Guess $\Sigma'$ with $|\Sigma'|\le q$ and mappings $\varrho$ and $\varphi$
for a label splitting $(\Sigma',E',\varrho,\varphi)$.
This takes $O(q+|E|)$ time, which is polynomial in the size of $TS$. To check if
the guess is correct, the at most $|S|^2$ SSPs of $TS$ need to be solved according to Lemma~\ref{l.embedssp}. An SSP is
represented by an inequality system polynomial in the size of $TS$, which in turn
can be solved by Karmarkar's algorithm\footnote{Karmarkar's algorithm finds a
rational solution of an inequality system (if it exists) in $O(|S|^{3.5}\cdot L^2\cdot \log L\cdot \log\log L)$
time with $L=|S|\cdot|\Sigma'|\cdot\log |c|$, where $c$ is the largest coefficient.
Our inequality systems do not contain constant terms, allowing for multiplication with
the common denominator to find an integer solution. The inequality system for an SSP
contains one true inequality relation ($\neq$), which can be avoided to apply
Karmarkar's algorithm by solving two separate systems, replacing $\neq$ by $<$ and $>$.
Alternatively, dual systems without $\neq$ may be solved in the rationals~\cite{BDS-SOFSEM-2016}.}
in polynomial time~\cite{karmarkar}.
\ENDBEW{s.optrelnp}

We now want to show that such a near-optimal label splitting is also \NP-hard.

\SATZ{s.optrelnphard}{Near-optimal Label Splitting is NP-hard}
For finite, reachable \lts $TS=(S,\Sigma,E,s_0)$ and $q\in\N$,
the near-optimal label splitting problem $(TS,q)$ is \NP-hard.
\ENDSATZ

To prove this theorem, we construct a polynomial-time reduction from some \NP-complete problem to
our problem. There are various \NP-complete problems on graphs known, but even when edge labels occur
in such problems, they are typically cost functions and the optimisation problem usually is to minimise the
cost for some walk. A direct translation to markings and regions does not look easy, especially since
we need to distinguish markings for an SSP and not minimise them. We chose to start our reduction at 
the problem of subset sums instead, which allows us to construct our own graph rather freely:

\DEF{d.subsetsum}{NP-Complete Subset Sum Problem~\cite{karp}\footnote{This problem is called \emph{Knapsack} in \cite{karp} and defined with
\(\Z\) instead of \(\np\), but the construction for the \NP-hardness proof 
produces non-negative numbers only. Without loss of generality, we may even assume that all
$c_i$ and $b$ are non-zero.
If some $c_i$ is zero, it can simply be excluded from $C$, and if $b=0$
the problem is trivially solvable (and we can replace it e.g.\ with 
$n=1$, $c_1=2$, $b=2$).}} 
Decide for $n\in\N$, $b\in\np$, and $C=\{c_1,\ldots,c_n\}\subseteq\np$, 
whether there is an index set $I\subseteq\{1,\ldots,n\}$ such that $\sum_{i\in I}c_i = b$.
\ENDDEF{d.subsetsum}

We make a reduction from the subset sum problem to our near-optimal label splitting problem.
For an input $(n,b,\{c_1,\ldots,c_n\})$ of a subset sum problem we determine two parameters: the unique
$k\in\N$ such that $2^k \le 1+2b+2\sum_{i=1}^nc_i < 2^{k+1}$ and $q=2n+k+11$. With the
parameter $k$, we construct the \lts shown in Fig.~\ref{f.optrellts}.
The value $q$ is the parameter for the label splitting problem. 

\begin{figure}[htbp]
\centering
\begin{tikzpicture}[LTS]
\draw(0,0) node[state,label=below left:$s_0$] (s0) {};
\path[edge] (s0) ++(-0.5,0) edge (s0);
\foreach \x in {1,2,3,4,5,6,7,8,9,10}
 \draw(\x,4) node[state] (u\x) {};
\draw[edge] (s0) edge node[auto]{$h_1$} (u1);
\foreach \x/\y/\u in {1/2/0,2/3/0,3/4/1,4/5/1,5/6/2,6/7/2,8/9/{k-1},9/10/{k-1}}
 \draw[edge] (u\x) edge node[auto]{$u_{\u}$} (u\y);
\foreach \x/\y/\u in {1/3/1,3/5/2,5/7/3,8/10/k}
 \draw[edge] (u\x) edge[bend right=20] node[auto,swap]{$u_{\u}$} (u\y);
\draw[dotted] (u7) edge (u9);

\foreach \x in {1,2,3,4,5,6,7}
 \draw(\x,2) node[state] (o\x) {};
\draw[edge] (s0) edge node[auto,swap,yshift=2mm]{$h_2$} (o1);
\draw[edge] (o1) edge (o2);
\draw[dashed] (o2) edge node[auto]{$u(1+2b+2\sum c_i)$} (o3);
\draw[edge] (o3) edge (o4);
\draw[edge] (o4) edge node[auto]{$o$} (o5);
\draw[dotted] (o5) edge node[auto,swap]{($n+1$ times)} (o6); 
\draw[edge] (o6) edge node[auto]{$o$} (o7);
\draw[edge] (o7) edge[bend right=50] node[auto,swap]{${\cal O}$} (o4);
\draw[edge] (o1) edge[bend right=20] node[auto,swap]{$o$} (o4);

\foreach \x in {1,2,3,4}
 \draw(\x,0.5) node[state] (sc\x) {};
\draw[edge] (s0) edge node[auto,swap,yshift=1mm]{$h_3$} (sc1);
\draw[edge] (sc1) edge (sc2);
\draw[dashed] (sc2) edge node[auto]{$u(\sum c_i)$} (sc3);
\draw[edge] (sc3) edge (sc4);
\draw[edge] (sc1) edge[bend right=20] node[auto,swap]{$\alpha$} (sc4);

\foreach \x in {1,2,3,4}
 \draw(\x,-1) node[state] (b\x) {};
\draw[edge] (s0) edge node[auto,yshift=-2mm]{$h_4$} (b1);
\draw[edge] (b1) edge (b2);
\draw[dashed] (b2) edge node[auto]{$u(2b)$} (b3);
\draw[edge] (b3) edge (b4);
\draw[edge] (b1) edge[bend right=20] node[auto,swap]{$\beta$} (b4);

\foreach \x in {1,2,3,4,5,6,7,8,9,10,11}
 \draw(\x,-3) node[state] (c\x) {};
\draw[edge] (s0) edge node[auto,near end,yshift=-2mm]{$h_5$} (c1);
\foreach \x/\y in {1/2,3/4,4/5,6/7,8/9,10/11}
 \draw[edge] (c\x) edge (c\y);
\foreach \x/\y/\u in {2/3/1,5/6/2,9/10/n}
 \draw[dashed] (c\x) edge node[auto]{$u(c_{\u})$} (c\y);
\draw[dotted] (c7) edge (c8);
\foreach \x/\y/\u in {1/4/1,4/7/2,8/11/n} {
 \draw[edge] (c\x) edge[bend right=20] node[auto,swap]{$\gamma_{\u}$} (c\y);
 \draw[edge] (c\y) edge[bend right=50] node[auto,swap]{$\gamma_{\u}$} (c\x);
}

\foreach \x in {1,2,3,4,5,6,7,8,9,10,11}
 \draw(\x,-4.5) node[state] (r\x) {};
\draw[edge] (s0) edge node[auto,swap]{$h_6$} (r1);
\foreach \x/\y/\u in {1/2/o,2/3/\alpha,3/4/o,4/5/{\gamma_1},5/6/o,6/7/{\gamma_2},8/9/o,9/10/{\gamma_n},10/11/{\cal {O}}}
 \draw[edge] (r\x) edge node[auto]{$\u$} (r\y);
\draw[dotted] (r7) edge (r8);
\draw[edge] (r1) edge[bend right=10] node[auto,swap]{$\beta$} (r11);
\end{tikzpicture}
\caption{The \lts constructed for a subset sum problem $(n,b,\{c_1,\ldots,c_n\})$
consists of six
strands starting with $h_1,\ldots,h_6$. The strands do not interact, they are only
connected by the fact that the same label must have the same effect (on a potential marking) even in different strands.
A term $u(x)$ denotes a binary encoding of some $x\in\N$ where each bit with value $2^j$
($0\le j\le k$) is expressed by the presence or absence of $u_j$ in the word $u(x)$.
The value of $k$ is chosen (just) high enough such that for every occurring 
value $x$ we get some valid binary encoding $u(x)$.
Dotted lines denote a canonical enumeration (up to $k$ or $n$ or $n+1$),
dashed lines, together with the edges before and after them, denote a sequence of
edges inscribed with some $u(x)$. The first four strands are auxiliary, they define
units of measurement $u_i$, assert distances via $o$ and ${\cal O}$, and allow to
express the two values $\sum_{i\in I}c_i$ and $b$ to be compared for the subset sum problem via $\alpha$ and $\beta$.
The strand $h_5$ will force the selection of some $c_i$ for an index set $I$ and strand $h_6$ will
assert the correct choice for the index set $I$, guaranteeing $\sum_{i\in I}c_i=b$
\label{f.optrellts}}
\end{figure}
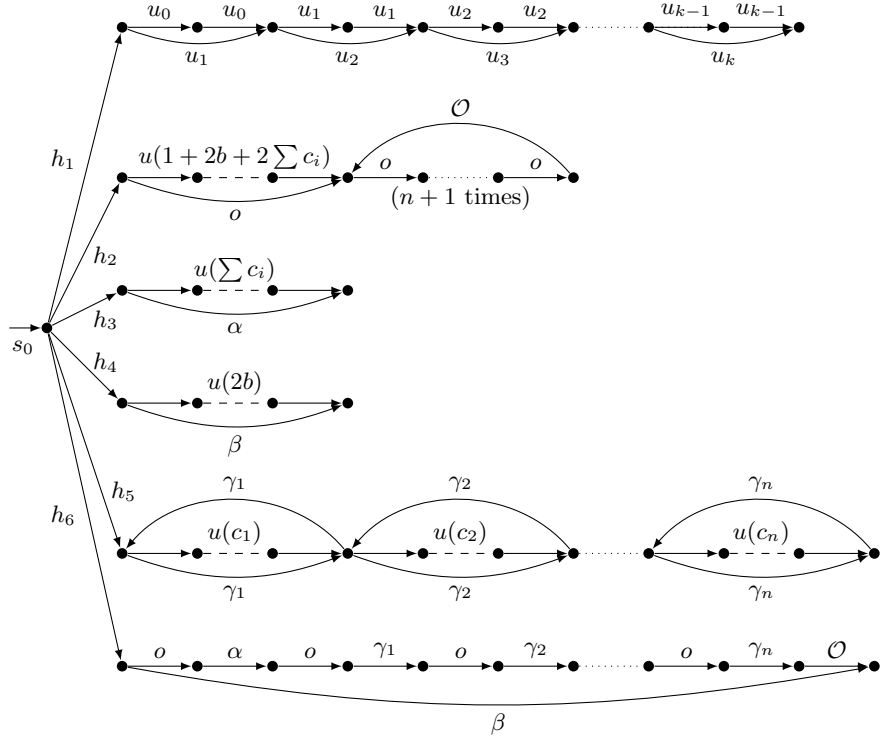

\section{Units and Region Values}
\label{sect.units_and_region_values} 

Let us assume a fixed subset sum problem with constructed parameters $k$, $q$, and the
\lts $TS=(S,\Sigma,E,s_0)$ from Fig.~\ref{f.optrellts}.
At the initial state $s_0$, there are six edges with labels $h_i$ ($1\le i\le 6$).
Everything following such an edge $h_i$ (but without the $h_i$-edge) will be called {\em the strand} $h_i$.
The strand $h_1$ defines a unit $u_0$ and some multiples, in the following sense:

\LEM{l.units}{Units}
Let $TS'$ be any reachable \lts embedding the strand $h_1$ of $TS$ in Fig.~\ref{f.optrellts}.
Let $r=(R,B,F)$ be a region of $TS'$ and let $s\step{v}s'$
by any walk in $TS'$ with $v\in\{u_0,u_1,\ldots,u_k\}^*$.
Then, $\E(v) = R(s')-R(s) = \sum_{i=0}^k \parikh(v)(u_i)\cdot 2^i\cdot\E(u_0)$.
\ENDLEM
\BEW
Let us name the states of strand $h_1$ as $s_i,s_i'$ for $1\le i\le k$ with a last state
$s_{k+1}$ such that $s_{i}\step{u_{i}}s_{i+1}$ (the lower, curved edges) and 
$s_i\step{u_{i-1}}s_i'\step{u_{i-1}}s_{i+1}$ (the upper edges) for $1\le i\le k$. Clearly, 
$\E(u_0) = F(u_0)-B(u_0) = R(s_1')-R(s_1)$. By definition of a region, edges with the same
label have the same effect, e.g. $R(s_2)-R(s_1')=\E(u_0)$. Thus, $R(s_2)-R(s_1)=2\E(u_0)=\E(u_1)$
as $s_1\step{u_1}s_2$, so $u_1$ has the effect $2\E(u_0)$. The same reasoning for
$s_2\step{u_{1}}s_2'\step{u_{1}}s_{3}$ and $s_2\step{u_2}s_3$ yields the effect $4\E(u_0)$ for $u_2$.
Recursively, we obtain $2^i\cdot\E(u_0)$ as the effect $\E(u_i)$. As the effects are added up over a walk,
$\E(v) = R(s')-R(s) = \sum_{i=0}^k \parikh(v)(u_i)\cdot 2^i\cdot\E(u_0)$.
\ENDBEW{l.units}

We define now a function allowing us to compute the effect on region values for most edges
in $TS$.

\DEF{d.units}{Unit Mapping}
Let $x\in\N$ with $x<2^{k+1}$. Let $x=\sum_{i=0}^km_i\cdot 2^i$ with $m_i\in\{0,1\}$ be
a binary encoding of $x$.
Define $u(x)=x_k\ldots x_1x_0\in\{u_0,\ldots,u_k\}^*$ to be the word with $x_i=u_i$ if $m_i=1$
and $x_i=\varepsilon$ if $m_i=0$.
\ENDDEF{d.units}

As an example, $u(25)=u_4u_3u_0$ since $25$ is written $11001$ as a binary number.
When we annotate a combination of an edge, a dashed line, and another edge in $TS$
by a word $u(x)$, this shall denote a sequence of states and edges where the edges
have the letters of $u(x)$ as labels. This occurs once in each strand $h_2$, $h_3$,
and $h_4$ (with $u(1+2b+2\sum_{i=0}^nc_i)$, $u(\sum_{i=1}^nc_i)$, and $u(2b)$ as inscription,
respectively), as well as $n$ times in the strand $h_5$ (with $u(c_i)$ for $1\le i\le n$).
We can now determine the effects on region values for most of the functional labels occurring in $TS$.

\LEM{l.rvals}{Region Values}
Let $TS'$ be any reachable \lts embedding the strands $h_1$ through $h_4$ of $TS$ in Fig.~\ref{f.optrellts}.
Let $r=(R,B,F)$ be a region of $TS'$ and $\E(u_0)$ be the effect of $u_0$.
Then, $\E(\alpha)=\sum_{i=1}^nc_i\cdot\E(u_0)$,
$\E(\beta)=2b\cdot\E(u_0)$, $\E(o)=(1+2b+2\sum_{i=1}^nc_i)\cdot\E(u_0)$,
and $\E({\cal O})=-(n+1)\cdot\E(o)$.
\ENDLEM
\BEW
As walks between the same two states yield the same effect on the region value,
the first part of strand $h_2$ enforces directly $\E(o)=(1+2b+2\sum_{i=1}^nc_i)\cdot\E(u_0)$
when applying Definition~\ref{d.units} and Lemma~\ref{l.units}.
\iffalse 
The second part reveals that $(n+1)$ labels $o$ followed by one ${\cal O}$ form a cycle,
thus have effect zero. Therefore, $\E({\cal O})+(n+1)\cdot\E(o)=0$.
In strand $h_3$, we obtain $\E(\alpha)=
\sum_{i=1}^nc_i\cdot\E(u_0)$.
For strand $h_4$ 
we get $\E(\beta)=
2b\cdot\E(u_0)$.
\else
The remaining parts of the lemma are derived analogously.
\fi
\ENDBEW{l.rvals}

%

Let us try to determine the region effect of the labels $\gamma_i$ in strand $h_5$ now.
We find that there are pairs of $\gamma_i$ forming cycles (asserting $\E(\gamma_i)=0$)
and $\gamma_i$ can also be expressed in units, yielding $\E(u_0)=0$. This collapses our
choice of regions and makes any pair of states from the same strand indistinguishable. 
The only remedy is to relabel one of the $\gamma_i$'s:

\LEM{l.gamma}{Required Label Splitting for $\gamma_i$}
Let $s,s'\in S$ be two states in strand $h_5$ of $TS$ such that $s\step{\gamma_i}s'\step{\gamma_i}s$
is fulfilled for some $i\in\N$. Then, the SSP $(s,s')$ is unsolvable. The only way to make this SSP solvable
is to relabel one of the two $\gamma_i$-edges via label splitting.
\ENDLEM
\BEW
The lower $\gamma_i$-label in strand $h_5$ lets us determine
$\E(\gamma_i)=c_i\cdot\E(u_0)$, while the upper $\gamma_i$-label yields
$\E(\gamma_i)+c_i\cdot\E(u_0)=0$ due to the cycle via $u(c_i)$. The only solution for both equations is
$\E(\gamma_i)=\E(u_0)=0$ (since we assumed $c_i\neq 0$ in the beginning). So, in every region $r=(R,B,F)$ of $TS$,
$R(s')=R(s)+\E(\gamma_i)=R(s)$, i.e.\ the two states cannot be distinguished.
Therefore, the SSP $(s,s')$ is unsolvable. 

Since no other edges are involved in the cycle $s\step{\gamma_i\gamma_i}s$, the SSP $(s,s')$ can only become
solvable by relabelling one (or both) of these two edges, giving the edges different labels. 
The only way to relabel an edge is via a label splitting.
Relabelling exactly one of the two $\gamma_i$-edges in strand $h_5$
to a new label $\overline{\gamma_i}$ will potentially allow regions with $\E(\gamma_i)\neq 0\neq \E(u_0)$.
\ENDBEW{l.gamma}

\section{Proving the polynomial-time reduction}
\label{sect.polynomial_time}

Let us count the number of different labels in $TS$, i.e. the alphabet size.
We have $k+1$ labels $u_i$, $n$ labels $\gamma_i$, and one each of:
$\alpha$, $\beta$, $o$, ${\cal O}$, $h_1$, $h_2$, $h_3$, $h_4$, $h_5$, $h_6$.
Summing these up, we come to $|\Sigma|=n+k+11$.
We will show now the following statements:
\begin{enumerate}
\item The construction of $TS$ can be done in polynomial time. 
\item If a label splitting applied to $TS$ introduces less than $q=2n+k+11$ labels, some SSP instances will be unsolvable.
 To be more precise, each label $\gamma_i$ ($1\le i\le n$) needs a new, ``opposite'' label $\overline{\gamma_i}$.
 This also means that all other labels must remain ``unsplit'', as we are not allowed to have more than $q$ labels.
\item There is at least one label splitting with exactly $q$ labels making all SSPs solvable if and only if
 the corresponding subset sum problem has a solution.
\end{enumerate}
This will show that our construction is a polynomial-time reduction, concluding the proof of
Theorem~\ref{s.optrelnphard}.

\LEM{l.polytime}{Polynomial-Time Construction}
The construction of $TS$ can be done in polynomial time (of the size of the input
subset sum problem $(n,b,\{c_1,\ldots,c_n\})$).
\ENDLEM
\BEW
Note first that the value of numbers occurring in the input, i.e. $b$ and the $c_i$, can be
exponentially larger than the size of the input in its binary encoding.
The function $u\colon\N\to\{u_0,\ldots,u_k\}^*$ can be directly applied to the binary
encoding of the input and uses only linear time.
Using this function in the strands $h_2$ to $h_6$
reduces the values $b$ and $c_i$ to a logarithmic number of states and edges, matching it with the
size of the input. The parameter $k$ is logarithmic in the value $1+2b+2\sum_{i=1}^nc_i$,
which is the largest occurring number.\footnote{For all occurring $u(x)$ in $TS$, $x$ is less or equal to this value.
}
Strands $h_2$ and $h_6$ contain sequences of states and edges of length $O(n)$, this
can be matched with the $n$ coefficients $c_i$ of the input. Overall, $TS$ has a size
linear in the input and can be constructed in polynomial time. 
\ENDBEW{l.polytime}

We now take a look at the conditions a label splitting must fulfill to make SSPs solvable.
We partition the SSPs into three sets, pairs of states in strand $h_6$, pairs in one of the
other five strands, and pairs from different strands (including the initial state).
We start with the latter case.

\LEM{l.strandssp}{Inter-strand SSPs are always Solvable}
Let $s,s'\in S$ be two states that do not lie in the same strand of $TS$.
Let $TS'$ be the result of some label splitting of $TS$. 
Then, the SSP $(s,s')$ is solvable in $TS'$.
\ENDLEM
\BEW
\iffalse
We can measure the maximal region value distance of two states in any single strand in units of $\E(u_0)$.
For $h_1$, let $s_1$ be the first and $s_1'$ the last state in the chain forming the strand, then
for any region $r=(R,B,F)$, $R(s_1')-R(s_1)=(2^{k+1}-1)\cdot\E(u_0)$ is this maximal distance.
For $h_2$, we get the value $(1+2b+2\sum_{i=1}^nc_i)\cdot(n+2)\cdot\E(u_0)$.
For the next three strands we obtain $\sum_{i=1}^nc_i\cdot\E(u_0)$, $2b\cdot\E(u_0)$,
and $\sum_{i=1}^nc_i\cdot\E(u_0)$.
For the strand $h_6$ we find an upper bound of $\mu:=(n+3)\cdot(1+2b+\sum_{i=1}^nc_i)\cdot\E(u_0)$,
adding up the values for $n+2$ $o$'s, the $\gamma_i$'s, and $\alpha$. 
Clearly, $\mu$ dominates all six maximal region value distances.

A region $r=(R,B,F)$ solving an SSP of $TS'$ is sufficiently defined by $R(s_0)$ and the effects $\E(t)$ for all labels $t$.
We can then set $F(t)=\E(t)$ and $B(t)=0$ (or $F(t)=0$ and $B(t)=-\E(t)$ in case $\E(t)<0$), 
not restricting the solutions of any SSP. The values $R(s)$ for states
except $s_0$ can be determined via the spanning tree. By setting $\E(h_i)=2i\cdot\mu$ for $1\le i\le 6$ and $R(s_0)=0$
we ensure that $R(s)\neq R(s')$, unless the only remaining free parameter $\E(u_0)$ is zero. Note that the effects
of edges with other labels are fully determined by $\E(u_0)$ according to Lemma~\ref{l.rvals} and~\ref{l.units}.
If $\E(u_0)=0$, we artificially set $\mu:=1$ to obtain the region value $2i$ for all states in strand $h_i$ and
zero for $s_0$. In all cases, we have found a region distinguishing $s$ and $s'$, solving the SSP $(s,s')$.
\else
Since the events \(h_i\) each appear only once in $TS$, we can without loss of generality assume that they still appear in $TS'$ and were not split/renamed.
We define a region \(r=(R,B,F)\) that assigns \(R(s'')=i\) to all states \(s''\) in strand \(h_i\) and \(R(s_0)=0\).
This already determines \(R\) and guarantees that SSP \((s,s')\) is solved.
To extend this into a region,
it suffices to set \(B(t)=0\) for all \(t\), \(F\) is \(F(h_i)=i\) for \(i=1,\dots,6\), and \(F(t)=0\) otherwise.

\(r\) satisfies the requirements of a region, because
\(B(t)=0 \le R(s)\in\N\) for all \(t\) and \(s\),
and the only events with non-zero effect each appear on a single edge and satisfy \(R(s')=R(s)+\E(t)\), thus this holds for all \(s\step{t}s'\).
\fi
\ENDBEW{l.strandssp}

\LEM{l.s5effect}{Effects in Strand Five}
Let $TS'$ be the result of some label splitting of
$TS$ that relabels exactly one instance of each $\gamma_i$
in strand $h_5$ and nothing else.
Let $r=(R,B,F)$ be a region of $TS'$ and $\E(u_0)$ be the effect of $u_0$.
Then, $\E(\gamma_i)=c_i\cdot\E(u_0)$ if the upper edge with label $\gamma_i$ was relabelled and otherwise $\E(\gamma_i)=-c_i\cdot\E(u_0)$. 
\ENDLEM
\BEW
Assume that the upper edge was relabelled. Then there are two walks $s\step{u(c_i)}s'$ and $s\step{\gamma_i}s'$ in strand five.
These two walks must have the same effect, so $\E(\gamma_i)=\E(u(c_i))=c_i\cdot\E(u_0)$
when applying Definition~\ref{d.units} and Lemma~\ref{l.units}.

In the other case, there is a cycle $s\step{u(c_i)\gamma_i}s$, thus $\E(\gamma_i)=-\E(u(c_i))=-c_i\cdot \E(u_0)$.
\ENDBEW{l.s5effect}

\LEM{l.s1to5}{SSPs in the First Five Strands}
Let $TS'$ be the result of some label splitting of
$TS$ that relabels exactly one instance of each $\gamma_i$
in strand $h_5$ and nothing else.
Assume a region $r=(R,B,F)$ of $TS'$ with $\E(u_0)\neq 0$. 
For every pair $(s,s')$ of states with $s\neq s'$ inside one
of the strands $h_i$ with $1\le i\le 5$ we obtain $R(s)\neq R(s')$.
\ENDLEM
\BEW
Since $s$ and $s'$ are in the same strand, there is a walk $v\in\{u_0,\ldots,u_k,o\}^*$ 
(cf. Fig.~\ref{f.optrellts} and Definition~\ref{d.units}) 
with either $s\step{v}s'$ or $s'\step{v}s$. W.l.o.g. assume $s\step{v}s'$. We compute
$R(s')=R(s)+\E(v)$ where the effect of each letter in $\{u_0,\ldots,u_k,o\}$ is a positive multiple of $\E(u_0)$
(see Lemma~\ref{l.units} and \ref{l.rvals}). 
Thus, we find some $m\in\np$ with $R(s')=R(s)+m\cdot\E(u_0)$ and due to $\E(u_0)\neq 0$, also
$R(s)\neq R(s')$ is true.
\ENDBEW{l.s1to5}

\LEM{l.s6}{SSPs in the Strand $h_6$}
Let $TS'$ be the result of some label splitting of
$TS$ that relabels exactly one instance of each $\gamma_i$
in strand $h_5$ and nothing else.
Assume a region $r=(R,B,F)$ of $TS'$ with $\E(u_0)\neq 0$. 
For every pair of states $(s,s')$ with $s\neq s'$ inside
the strand $h_6$ we get $R(s)\neq R(s')$.
\ENDLEM
\BEW
Without loss of generality, assume that $\E(u_0)>0$ (otherwise negate all values to obtain a region ``$-r$'').
Note that $\E(o) > 2\E(\alpha)+\E(\beta)$
by Lemma~\ref{l.rvals}, 
as well as $\E(o) > 2\E(\gamma_i)$ for $1\le i\le n$
by Lemma~\ref{l.s5effect}. 
In any walk $s_1\step{x}s_2\step{o}s_3\step{y}s_4$ with $x,y\in\{\alpha,\gamma_1,\ldots,\gamma_n\}$,
the difference $R(s_3)-R(s_2)=\E(o)$ is so big that all four states must have pairwise different region values,
with $R(s_3)$ and $R(s_4)$ being greater than both $R(s_1)$ and $R(s_2)$. The same holds with
$s_2\step{\beta}s_1$ and $s_2\step{o}s_3\step{\alpha}s_4$ and also for the case
$s_4\step{\gamma_n}s_3\step{{\cal O}}s_2$ and $s_1\step{\beta}s_2$ where $\E({\cal O})<-\E(o)$.
If we order the states of strand $h_6$ into pairs connected by an edge label from $\{\alpha,\beta,\gamma_1,\ldots,\gamma_n\}$,
the pair at $\beta$ has the lowest region values, and the values increase through the $\alpha$ and $\gamma_1$ pairs
up to the states adjacent to $\gamma_n$. The two states inside a pair also have different values since $\E(u_0)\neq 0$
(see Lemmas~\ref{l.rvals} and~\ref{l.s5effect}). 
\ENDBEW{l.s6}

To be a region, a triple of mappings $r=(R,B,F)$ must fulfill the two conditions $R(s)\ge B(t)$ and
$R(s')=R(s)-B(t)+F(t)$ for every edge $s\step{t}s'$. If $R(s)<B(t)$ for some $s\step{t}$, we can determine
$c:=B(t)-R(s)$ and modify the mappings to $r'=(R+c,B,F)$. The new triple $r'$ distinguishes the same
pairs of states as $r$, so it solves the same SSPs. 
We can easily see that the second condition holds in the first five strands
if $\E$ satisfies the requirements of Lemmas~\ref{l.units}, \ref{l.rvals}, and \ref{l.s5effect}. 

\LEM{l.gcycles}{$u_0$ provides Region Effects in the First Five Strands}
Take only the first five strands of $TS$, apply a label splitting that
relabels exactly one instance of each $\gamma_i$ in strand $h_5$ (and
nothing else), and call the resulting \lts $TS'$.
Let $\E$ be a mapping $\E\colon\Sigma\to\Z$ that fulfills the equations given in Lemma~\ref{l.units},
\ref{l.rvals}, and \ref{l.s5effect}.
There is a region $(R,B,F)$ of $TS'$ such that $F-B=\E$.
\ENDLEM
\BEW
\iffalse
Just observe that in every cycle in one of the first four strands, there is one edge whose defining
value $\E(t)=F(t)-B(t)$ is determined from the other edges of the cycle. In the equation 
$R(s')=R(s)-B(t)+F(t)=R(s)+\E(t)$ only this difference $\E(t)$ is important.
In strand $h_3$ e.g.,
$\E(\alpha)$ (and thus $B(\alpha)$ and $F(\alpha)$) can be computed from the $u_i$-edges. The effect of
every label from $\{u_0,\ldots,u_k,o,{\cal O},\alpha,\beta\}$ is defined only once in this way.
After the label splitting, we obtain analogous findings for each $\gamma_i$ and $\overline{\gamma_i}$ (the relabelled
version of $\gamma_i$) in strand $h_5$. For one label of each pair, the cycle $u(c_i)\gamma_i$ or $u(c_i)\overline{\gamma_i}$
defines the value of $\E(\gamma_i)$ or $\E(\overline{\gamma_i})$, the other value is then obtained from
the cycle $\gamma_i\overline{\gamma_i}$.
%
\else
Assume an arbitrary, fixed spanning tree of $TS'$. Remember that this assigns a unique walk $s_0\step{\sigma_s}s$ to each state $s$.
Define \(R(s_0)=\max\lbrace 0\rbrace\cup\lbrace -\E(\sigma_s)\mid s\in S \rbrace\)
and extend this via \(R(s)=R(s_0)+\E(\sigma_s)\) to all states.
The choice of \(R(s_0)\) guarantees \(R(s)\geq 0\) for all states.
Next, define \(F(t)=\E(t)\) and \(B(t)=0\) for all \(t\) with \(\E(t)\geq 0\), and \(F(t)=0\) and \(B(t)=-\E(t)\) otherwise.
We can now see that \((R,B,F)\) is a region, i.e.\ that for all \(s\step{t}s'\) we have \(R(s)\geq B(t)\) and \(R(s')=R(s)+\E(t)\).
Both conditions follow trivially for edges in the spanning tree. Assume \(s\step{t}s'\) to be a chord.
Every chord completes a (generalised) cycle, therefore the effects defined via
such cycles in Lemmas~\ref{l.units}, \ref{l.rvals}, and~\ref{l.s5effect} ensure $R(s')=R(s)+\E(t)$.
\fi
\ENDBEW{l.gcycles}

As a consequence, all SSPs becoming solvable depends on only two points now: That we use a label splitting
that relabels one of each $\gamma_i$ for $1\le i\le n$ in strand $h_5$ and that the two walks in strand
$h_6$ from the leftmost to the rightmost state have the same region effect, i.e. that the (generalised) cycle 
formed by these two walks has effect zero. We will see next that this is equivalent to finding a solution to the subset sum
problem.

\SATZ{l.equiv}{Subset Sum Solution is Equivalent to Solvable SSPs}
Let ${\cal S}=(n,b,\{c_1,\ldots,c_n\})$ be a subset sum problem and $TS$ (with the computed $k\in\N$)
be the \lts constructed from it as per Figure~\ref{f.optrellts}.
${\cal S}$ has a solution if and only if there is a PN-embeddable \lts $TS'$ (i.e.\ all SSPs are solvable)  
resulting from $TS$ by a label splitting with at most $q=2n+k+11$ labels.
\ENDSATZ
\BEW
Let $I$ be an index set of our subset sum problem ${\cal S}$ (not necessarily a solution, though).
Define a label splitting $(\Sigma',E',\varrho,\varphi)$ with $\Sigma'=\Sigma\cup\{\overline{\gamma_i}\mid 1\le i\le n\}$,
$\varrho(t)=t$ for $t\in\Sigma$, and $\varrho(\overline{\gamma_i})=\gamma_i$ for $1\le i\le n$.
Relabel in strand $h_5$ of $TS$ the upper $\gamma_i$ to $\overline{\gamma_i}$ if $i\in I$,
and relabel the lower $\gamma_i$ to $\overline{\gamma_i}$ if $i\notin I$.
Note how this changes the region
effect of $\gamma_i$: If $i\in I$, $\E(\gamma_i)=c_i\cdot\E(u_0)$, and if $i\notin I$,
then $\E(\gamma_i)=-c_i\cdot\E(u_0)$
(see Lemma~\ref{l.s5effect}). 
The relabelled \lts has now exactly $q=2n+k+11$ different labels. 

Let us take a look at the strand $h_6$ now. The upper walk from the leftmost to the rightmost state contains
exactly $n+1$ $o$'s and one ${\cal O}$. By Lemma~\ref{l.rvals}, $\E({\cal O})=-(n+1)\cdot\E(o)$,
thus the region effects of these labels cancel each other out. The remaining effect of the upper walk is
then $\E(\alpha\gamma_1\ldots\gamma_n)$ with $\E(\alpha)=\sum_{i=1}^nc_i\cdot\E(u_0)$.
Therefore, if $\E(\gamma_i)=-c_i\cdot\E(u_0)$ it will cancel out the $c_i\cdot\E(u_0)$ in $\E(\alpha)$,
while a positive $\E(\gamma_i)=c_i\cdot\E(u_0)$ will double the effect.
Overall, we get $\E(\alpha\gamma_1\ldots\gamma_n)=2\sum_{i\in I}c_i\cdot\E(u_0)$.

If $I$ is a solution of the subset sum problem, we have $\sum_{i\in I}c_i=b$ and we obtain
$\E(\alpha\gamma_1\ldots\gamma_n)=2\sum_{i\in I}c_i\cdot\E(u_0)=2b\cdot\E(u_0)=\E(\beta)$.
\iffalse
By defining $R(s_0)=0$ and $\E(u_0)=1$ we can build a triple $r=(R,B,F)$ with $B(t)=0$ and
$F(t)=\E(t)$ (or $B(t)=-\E(t)$ and $F(t)=0$ in case $\E(t)<0$) for the functional labels $t$
and the effects of $h_1$ through $h_6$ set according to the proof of Lemma~\ref{l.strandssp}.
Immediately, $R(s)\ge B(t)$ is guaranteed for any edge $s\step{t}s'$, and for the first five
strands, Lemma~\ref{l.gcycles} shows $R(s')=R(s)-B(t)+F(t)$. 
\else
By defining \(\E(u_0)=1\) and \(\E(h_i)=0\) and extending this to a mapping \(\Sigma\to\Z\) via the requirements from
Lemma~\ref{l.units} (for $u_i$),
Lemma~\ref{l.rvals} (for $\alpha$, $\beta$, $o$, and ${\cal O}$), and
Lemma~\ref{l.s5effect} (for $\gamma_i$),
the preconditions of Lemma~\ref{l.gcycles} are fulfilled. This provides us with a region $r$ for our effects $\E$.
\fi
For strand $h_6$, we can derive
the effect of $s\step{\beta}s'$ via the upper chain of edges as
$R(s')=R(s)+\E(\alpha\gamma_1\ldots\gamma_n)=R(s)+\E(\beta)$ as required.
Thus, $r=(R,B,F)$ is a region with $\E(u_0)\neq 0$, and according to the
Lemmas~\ref{l.s1to5}, and~\ref{l.s6} it solves all SSPs inside of the same strand. SSPs between strands are always solvable by Lemma~\ref{l.strandssp}.
We conclude that the \lts is PN-embeddable.
 

Assume for the other direction that we find a label splitting with at most $q=2n+k+11$ different
labels that leads to a PN-embeddable, relabelled \lts, i.e. all SSPs are solvable.
Any label splitting yielding less than $q$
different labels has a guaranteed unsolvable SSP by Lemma~\ref{l.gamma}. Therefore, the alphabet
size is exactly $q$, which means that one of each pair of $\gamma_i$ has been relabelled, and nothing
else.\footnote{We exclude simple renamings without loss of generality.}
Thus, we have used a label splitting
as constructed in the first paragraph of this proof, which stems from some arbitrary index set $I$.
Let $s,s'$ be the first two states in strand $h_1$, with $s\step{u_0}s'$. There is a region
$r=(R,B,F)$ with $R(s')-R(s)=\E(u_0)\neq 0$ solving the SSP $(s,s')$. Strand $h_2$ provides again
$\E({\cal O})=-(n+1)\cdot\E(o)$, so that in strand $h_6$ we find
$2\sum_{i\in I}c_i\cdot\E(u_0)=\E(\alpha\gamma_1\ldots\gamma_n)=\E(\beta)=2b\cdot\E(u_0)$.
Dividing by $2\E(u_0)$ we get $\sum_{i\in I}c_i=b$, therefore the index set $I$ is a solution
of the subset sum problem.
\ENDBEW{l.equiv}

\section{Concluding remarks}\label{sect.O}

Synthesising a small model like a Petri net from a large \lts or a set of
observable processes can be done in more than one way. For Process Mining~\cite{vdaalst},
Badouel and Schlachter~\cite{bs17} have constructed an incremental
over-approximation algorithm. This allows behaviour that has not been observed
essentially by adding edges to the LTS.
Carmona~\cite{carmona-label-splitting} introduced a heuristic
label splitting algorithm that can relabel a finite \lts to make it
the reachability graph of a Petri net. 
For neither of the two approaches the time complexity is known. Both
could well be exponential. A first polynomial-time label splitting algorithm
has been shown in~\cite{sw19}, but it will not always generate optimal results.

In this paper, we have investigated a case where both over-approximation 
(by embedding into a Petri net reachability graph) and
label splitting are allowed in the synthesis procedure.
Finding a Petri net with a minimal alphabet size or even only limiting the
alphabet size makes this problem \NP-complete, therefore a small model is
not easily obtainable. The \NP-completeness proof could be modified in certain
ways, e.g. we can ask whether the removal of a certain number of edges (instead
of label splitting) can make an \lts embeddable into a Petri net reachability
graph. This leads to a model with a fault tolerance, i.e. a few desired behaviours may
be missing and additional ones will exist at the same time. A problem instance 
would be $(TS,n)$, where $n$ edges may be removed from an \lts $TS$ to make it PN-embeddable.
The construction for our reduction remains the same, but we 
would need to remove one of each pair of
$\gamma_i$-edges in strand $h_5$ now to allow for a solution (instead of relabelling them). 
The remainder of the proof will stay the same, so this problem
is also \NP-complete.

We believe that label splitting aiming at exact synthesis, making
the \lts isomorphic to a Petri net reachability graph, should not be easier than
embedding, since the exact synthesis additionally demands all ESSPs to be solvable.
We have no proof for this conjecture, though. In our construction, a label $x$
in any of the six strands of $TS$ has the effect $\E(x)=m\cdot\E(u_0)$ for some
$m\in\N$ and the unit label $u_0$. If an edge $s\step{x}s'$ occurs, the solvability
of all ESSPs demands that also $s\step{u_0^m}s'$ must be possible. Here, $m$ has
essentially the same size as the parameter values of the input, a subset sum problem.
As the subset sum problem can be written in a binary encoding, 
our constructed \lts $TS$ with $m$ $u_0$-edges will be exponential in the size of the input.
Therefore, our construction cannot be done in polynomial time anymore.

%
Finally, observe that the embebbing problem for unlabelled Petri nets is a special
case of our problem, where the parameter $q$ for an input instance $(TS,q)$ must be set
so that no label splitting can occur. Our reduction only works for trivial instances
of the subset sum problem in this case, and indeed the unlabelled embedding problem
is a subproblem of the exact synthesis problem for unlabelled nets, solvable with 
a polynomial-time algorithm.

\bibliographystyle{plain}

\end{document}